\documentclass[]{interact}

\usepackage{epstopdf}
\usepackage{subfigure}
\usepackage[title]{appendix}
\usepackage{natbib}
\usepackage{booktabs}
\usepackage{lscape}
\usepackage{hyperref}
\usepackage[all]{hypcap}
\usepackage{mwe} 
\hypersetup{hypertex=true,
colorlinks=true,
linkcolor=blue,
anchorcolor=blue,
citecolor=blue}

\bibpunct[, ]{(}{)}{;}{a}{}{,}
\renewcommand\bibfont{\fontsize{10}{12}\selectfont}

\theoremstyle{plain}

\theoremstyle{definition}

\theoremstyle{remark}

\begin{document}

\articletype{ARTICLE}

\title{The Impact of Customer Online Satisfaction on Stock Returns: Evidence from the E-commerce Reviews in China}

\author{
\name{Zhi Su\textsuperscript{a}, Danni Wu\textsuperscript{a}, Zhenkun Zhou\textsuperscript{b,c}, Junran Wu\textsuperscript{d} and Libo Yin\textsuperscript{e,*}\thanks{*Corresponding author: yinlibowsxbb@126.com}}
\affil{
\textsuperscript{a}School of Statistics and Mathematics, Central University of Finance and Economics, Beijing, China\\
\textsuperscript{b}School of Statistics, Capital University of Economics and Business, Beijing, China\\
\textsuperscript{c}China Consumption Big Data Academy, Capital University of Economics and Business, Beijing, China\\
\textsuperscript{d}State Key Lab of Software Development Environment, Beihang University, Beijing, China\\
\textsuperscript{e}School of Finance, Central University of Finance and Economics, Beijing, China} 
}

\maketitle

\begin{abstract}
This paper investigates the significance of consumer opinions in relation to value in China's A-share market. By analyzing a large dataset comprising over 18 million product reviews by customers on JD.com, we demonstrate that sentiments expressed in consumer reviews can influence stock returns, indicating that consumer opinions contain valuable information that can impact the stock market. Our findings show that Customer Negative Sentiment Tendency (CNST) and One-Star Tendency (OST) have a negative effect on expected stock returns, even after controlling for firm characteristics such as market risk, illiquidity, idiosyncratic volatility, and asset growth. Further analysis reveals that the predictive power of CNST is stronger in firms with high sentiment conditions, growth companies, and firms with lower accounting transparency. We also find that CNST negatively predicts revenue surprises, earnings surprises, and cash flow shocks. These results suggest that online satisfaction derived from big data analysis of customer reviews contains novel information about firms’ fundamentals.

\end{abstract}

\begin{keywords}
Consumer opinions; Stock return; China’s A-share market; Cash flow surprises
\end{keywords}

\section{Introduction}

Customer satisfaction is a critical factor in a company’s success, as cash flow generation largely depends on the value provided to customers \citep{agarwal2021disaggregated}. Despite this, there is limited research on the extent to which consumer opinions can offer insights into a company’s future cash flows and stock returns \citep{huang2018customer}. With the growth of e-commerce platforms, investors have access to vast amounts of data on consumer online behavior. However, few studies have examined the impact of consumer information on stock return in the Chinese market due to data availability limitations. This study aims to close this research gap by examining the effect of online reviews and consumer attitudes on the Chinese stock market. 

By analyzing and mining alternative data, this paper examines the information content and investment value of such data, enriching relevant research. Alternative data refers to real-time and detailed information that differs from conventional financial reports and filings obtained from sources such as investor presentations, analyst reports, and company filings \citep{monk2019rethinking}. The ability of alternative data to provide excess returns and forward-looking information is a concern for both investors and researchers. Past research has shown that online customer activity data can provide insight into a firm’s fundamentals, enabling anticipation of standardized unexpected earnings, revenue, and announcement returns \citep{froot2017measures,zhu2019big}. Our work provides empirical evidence that alternative data, such as consumer reviews data, can be used to investigate the impact of consumer satisfaction on stock returns. Furthermore, we demonstrate that consumer reviews affecting stock returns are not entirely consistent with traditional financial information and contain additional information beyond financial data.

This paper explores the investment value of consumer opinions using alternative data and measuring such opinions through customer product reviews posted on JD.com, a prominent online retailer in China. By analyzing over 18 million consumer product reviews between 2008 and 2017, we have gathered compelling evidence supporting the notion that consumer opinions hold investment value. We construct static and dynamic panel models with a vector of firm characteristics, and results show that future stock returns are negatively correlated with both Customer Negative Sentiment Tendency (CNST) and One-Star Tendency (OST). The conclusions investigate the impact of consumer online satisfaction stock returns in China’s A-share market.

By using the Consumer Confidence Index (CCIs) to evaluate sentiment in the Chinese stock market, we can analyze the impact of future stock returns under various economic conditions and present our research findings. CNST can significantly and negatively predict future stock returns during high sentiment, while low sentiment has little predictive potential for future stock returns. This negative impact appears to be concentrated for different types of companies, and we use return on assets(ROA) as a measure of company profitability, book-to-market(B/M) as a measure of growth or value companies, and asset growth(AG) as a measure of company investment. CNST can significantly negatively predict future stock returns during high ROA, high asset growth, and low book-to-market. Additionally, CNST can significantly negatively predict future stock returns during high earnings aggressiveness (EA) and earnings smoothing (ES). Based on these heterogeneity tests, we also analyze the reasons why e-commerce sales data can be used for prediction. CNST can also negatively predict revenue surprises, earnings surprises, and cash flow shocks. The results indicate that consumer online reviews and opinions contain valuable insights regarding future cash flows that were previously unrecognized and future stock returns.

Our study makes the following contributions to the literature. 
First, our research contributes to understanding the influence of alternative data in financial markets by examining the effects of online review information on stock returns. Despite the recent surge in research exploring the impact of consumer behavior data on financial markets, previous findings have overlooked the influence of consumer opinions regarding product emotions. We present the first formal empirical evidence confirming that consumer opinions have a negative impact on expected stock returns and that these opinions are composed of features extracted using a tendentious lexicon.
Second, we consider a range of established predictors in the panel analysis of cash flow, as well as factors likely to be associated with online consumer opinions. Our results indicate that aggregated customer reviews provide valuable new insights not obtainable from traditional sources such as financial statements. Additionally, results also indicate that customer online opinions may offer new insights into future revenue cash flows. We pay attention to a novel perspective of consumer companies, that is, the consumer online opinion of consumer companies. Given the substantial increase in data generated by large crowds, it would be worthwhile for future research to explore how this information can be leveraged in other corporate scenarios.
Third, this study sheds light on the factors contributing to the predictability of stock returns based on consumer opinions. In particular, it examines whether predictability is especially prominent for stocks experiencing periods of high sentiment, lower accounting transparency of corporate information, and growth companies.

The remainder of this paper is organized as follows. Section \ref{sec:lit} discusses the related literature. Section \ref{sec:data}  describes the data and variable definitions. Section \ref{sec:res} presents the empirical results, and Section \ref{sec:con} concludes.

\section{Literature review}
\label{sec:lit}
As far as we know, this study is relevant to the fourth research direction of literature: consumer-related information, product-related information, the wisdom of crowds, and customer capital.

First, due to the lack of data related to consumer information in early studies, most literature has mainly studied the influence of consumer satisfaction on firm performance and stock valuation through customer survey data \citep{anderson2004customer,ittner2009commentary,jacobson2009financial,luo2010customer,tuli2009customer}
Previous research has generally established a positive correlation between higher consumer satisfaction\footnote{The customer satisfaction data comes from the American Customer Satisfaction Index (www.theacsi.org).}, improved future operating performance, and higher firm valuation. The emergence of technologies such as big data and cloud computing has enabled the utilization of consumer behavior data in academic research \citep{da2011search,froot2017measures,huang2018customer,zhu2019big,agarwal2021disaggregated,li2022can}
For example, \citet{huang2018customer} have shown that customer ratings on Amazon.com can be used to predict a company’s future performance and stock returns but did not combine consumers’ opinions on product emotions. \citet{li2022can} examines the impact of online sales disclosure on earnings management for companies in consumer industries. The results indicate that earnings management decreases when third-party online sales data is publicly disclosed in a well-known financial database.

Second, previous empirical studies have revealed that product recalls are linked to a negative abnormal return on the stock price. This suggests that negative impacts on brand equity, company reputation, inferior product quality, and poor customer perception can lead to a negative impact on stock returns \citep{hoffer1988impact,barber1996product,zhao2013financial}
Recent research by \citet{da2011searchfun} demonstrates that the Search Volume Index (SVI) of a company’s products on Google can provide insight into the company’s future fundamentals. This information can be used to predict revenue and earnings surprises, as well as earnings announcement returns. Additionally, the SVI is a valuable leading indicator for a firm’s future cash flow and is relevant for determining the firm’s value.  \citet{tang2018wisdom} find that information about products created by third parties and shared on Twitter can forecast a company’s fundamentals and performance. Our study contributes to this body of research by focusing on a method that provides a direct measurement of how customers feel about the quality of a product.

Third, with the rise of online platforms, users have accumulated large groups of digital records of historical behaviors on various platforms, such as comments and sharing opinions on investment and social media platforms. Therefore, aggregating the behavior records of a large number of users may reflect some behavioral patterns of the group. There is an increasing body of literature that delves into the influence of behavioral information aggregation and the wisdom of crowds on stock prices. For example, previous research shows that a significant number of financial market participants conveyed information about future stock returns and cash flows through their collective actions \citep{kelley2013wise,chen2014wisdom,lee2015search}.
\citet{green2019crowdsourced}, \citet{sheng2022asset}, \citet{chi2021employee}, and \citet{symitsi2021employee} used millions of employee online reviews data from the Glassdoor website, the largest website for employee reviews, and found that the summarized company evaluation value can generate investment-relevant signals and significantly predict stock market reactions. Different from the above literature, our paper uses the aggregation of consumers’ online shopping information. This is due to the fact that customers are among the most significant nonfinancial stakeholders of companies.

Fourth, there is a growing body of literature on the relationship between customer capital and finance, which refers to customers’ brand loyalty toward a firm. The interplay between a firm’s financial and product market attributes was first explored by \citet{titman1984effect} and \citet{titman1988determinants}, both theoretically and empirically. Since then, several studies have investigated the impact of financial attributes on a firm’s decisions and performance in the product market \citep{fresard2010financial,phillips2013firm,gourio2014customer,gilchrist2017inflation}.
Likewise, various studies have examined the effect of product market attributes on corporate policies and valuation\citep{belo2014brand,vitorino2014understanding,gourio2014customer,dou2021external}.
For example, \citet{dou2021inalienable} proposed the Inalienability of Customer Capital (ICC) model, which explores the relationship between customer capital and key talents’ contributions and pure brand recognition. Their findings suggest that firms with higher ICC have greater average returns, higher talent turnover, and more cautious financial policies. Our study contributes to this literature by analyzing customer product reviews and deriving two categories of
features to measure customer satisfaction with products and brands.

\section{Data and variable definitions}
\label{sec:data}
This section provides an overview of the key variables and describes the data sources used in our analysis.

\subsection{Online consumer review data}

JD.com is one of the top online retailers in China, with sales reaching a staggering 672 billion dollars in 2018. The company was established in June 1998, but it did not begin allowing consumers to publish reviews of its products until 2008 when it finished developing its 3C (Computer, Communication, and Consumer Electronics) division. Since its inception, over 305 million customers have made purchases on the website and left over one billion reviews, establishing it as the foremost provider of consumer reviews on the Internet. According to the guidelines for creating reviews provided by JD.com, anyone who has purchased a product through JD.com is eligible to write a review of that product. Reviews should concentrate on particular aspects of the product, the quality of the service received, and consumer (or merchant) reply content in transaction orders. Thus, it is likely that consumers' judgments of a product's quality and value are reflected in the reviews provided by customers on JD.com. The criteria forbid unreasonable reviews, hostile reviews, and useless reviews in an effort to reduce the number of potential conflicts of interest. In addition, JD.com uses complex technical measures, such as a captcha, to protect its website from being hacked by spam bots and other malicious software. As a result, the customer reviews that can be seen on JD.com are both trustworthy and of high quality. The reviews encompass not only the product but also the seller's performance, such as the shipping experience, packaging, and product availability. Additionally, any factors that could potentially impact an investor's decision, including a customer's impression of the brand, are also taken into consideration. The feedback provided by customers typically consists of a rating system ranging from one to five stars, with one representing the lowest rating. Additionally, customers are given the opportunity to include a text review. The posting time of each review is recorded, enabling the tracking of consumer opinions over a period of time.

To ensure a comprehensive selection of consumer reviews, we source the list of companies from Sina Finance, one of the largest financial news portals in China. Sina Finance categorizes the entire stock market into various sectors based on stock characteristics. Following our previous work \citep{wu2020predicting}, we select four sectors most relevant to consumer products: home appliances, garments, wine, and food industries. Within these four sectors, there are a total of 177 companies with market values ranging from billions to hundreds of billions, representing diverse market capitalization firms. To obtain online customer review data, we used stock names and codes as seeds. We manually searched for company names on JD.com to identify public firms with customer product reviews and checked whether these firms sell products on the website. We retrieved the list of brands from JD.com and identified the firms that own these brands. In total, we ultimately found 106 businesses with JD.com customer product reviews.

We created a web crawler to collect reviews for a sample of public firms, using each brand as a search term on JD.com and searching all reviews for products. We then gathered information about each product, including the product’s name, the brand’s name, the date of review, star rating in numbers, review text, and the number of days between the order and review dates. The time period covered by the sample of reviews ranges from 2008 to 2017, and we also deleted duplicate reviews made on the same day for the same product by the same reviewer account ID. We created a panel dataset consisting of firm-weeks that include customer reviews. To ensure data reliability, we set a threshold of at least 1000 online reviews and a minimum time span of 12 months. for each firm. The final sample has 21,703 firm-weeks. Table \ref{tab:sumreview} provides information on the number of reviews, products, and firms in the entire sample, as well as for each of the four industry sectors. There are almost 18 million reviews available online, covering 164,715 different goods created by the sample companies. The home appliance sector and the garment industry have the highest number of product reviews, with more than 7.2 million and 5.3 million reviews respectively. Together, these two industries account for 70 percent of the reviews we have acquired.

~\\
\centerline{[Insert Table \ref{tab:sumreview} here]}%

\subsection{Variables}
\label{section:var}

\subsubsection{Main Variables}

According to the existing literature \citep{luo2009quantifying,tirunillai2012does,luo2013social,huang2018customer,zhou2018tales}, we derived two categories of features. The first category consists of characteristics extracted from a review based on tendentious words in the review text. These words reflect specific attitudes towards the purchasing experience, such as positive or negative sentiment, and their correlation with stock returns varies based on the type of attitude expressed. The second category is based on the review rating, ranging from one to five, which indicates the level of customer satisfaction with the purchase experience. This category complements the first by providing an additional measure of satisfaction.

~\\
\textbf{Customer (Positive/Negative) Sentiment Tendency}
~\\

The \textit{Customer Sentiment Tendency} category is based on tendentious words extracted from review text on JD.com, a platform where consumers share their experiences with products and services, generating a large amount of unstructured data that reflects their emotions and attitudes. Furthermore, the relationship between various tendency reviews and stock return varies \citep{luo2009quantifying,tirunillai2012does,zhou2018tales}. Therefore, we construct the indicators of negative and positive emotional trends. We define the Customer (Positive/Negative) Sentiment Tendency of firm $i$ in week $t$ as 

\begin{equation}
N(\mathrm{rev})^{neg}_{i,t}=\sum_r^N\{1\ if\ NW_{i,t,r}\ >\ PW_{i,t,r}\ else\ 0\}
\label{equ:neg}
\end{equation}

\begin{equation}
N(\mathrm{rev})^{pos}_{i,t}=\sum_r^N\{1\ if\ PW_{i,t,r}\ >\ NW_{i,t,r}\ else\ 0\}
\label{equ:pos}
\end{equation}

\begin{equation}
Diff(\mathrm{rev})^{neg}_{i,t}=N(rev)^{neg}_{i,t}-N(\mathrm{rev})^{neg}_{i,t-1}
\label{equ:neg2}
\end{equation}

\begin{equation}
Diff(\mathrm{rev})^{pos}_{i,t}=N(\mathrm{rev})^{pos}_{i,t}-N(\mathrm{rev})^{pos}_{i,t-1}
\label{equ:pos2}
\end{equation}

where $N$ represents the number of reviews published in $n$ weeks, with $n$ =1,2\ldots,12. $NW_{i,t,r}$ and $PW_{i,t,r}$represent the counts of negative and positive words, respectively, in review $r$ for firm $i$ in week $t$. If the count of negative words in a review is higher than that of positive words, then $N(\mathrm{rev})^{neg}_{i,t}$ is equal to 1. Similarly, if the count of positive words is higher than that of negative words, then $N(\mathrm{rev})^{pos}_{i,t}$ is equal to 1. $Diff(\mathrm{rev})^{neg}_{i,t}$ and $Diff(\mathrm{rev})^{pos}_{i,t}$ represent the differences between week $t$ and week $t-1$. Customer Positive Sentiment Tendency (CPST) typically involves favorable emotions and suggestions to purchase particular products. In contrast, Customer Negative Sentiment Tendency (CNST) relates to negative emotions and recommendations not to purchase particular items. This includes derogatory comments about the brand, criticism of the product itself, and complaints regarding instances of product malfunctions or poor customer service that led to unsatisfactory experiences. Table \ref{tab:sum} shows that the mean of $Diff(\mathrm{rev})^{neg}_{i,t}$ is approximately zero, while the mean of $Diff(\mathrm{rev})^{pos}_{i,t}$ is 0.01.

~\\
\textbf{Star Tendency (Five-Star Tendency / One-Star Tendency)}
~\\

The \textit{Star Tendency} category is derived from the numerical review ratings that customers have given the product. These ratings provide a clear indication of how customers feel about the product’s quality, value, and overall experience. Earlier studies have shown that customer review ratings have a significant relationship with stock returns \citep{huang2018customer,luo2013social}. To account for potential biases in product ratings on JD.com, we have created two distinct categories for Star Tendency: Five-Star Tendency (FST) and One-Star Tendency (OST). Products on JD.com are evaluated on a scale of one to five stars, where one corresponds to the lowest rating and five represents the highest. However, there is a possibility of selection bias on JD.com, where products sold on the platform may be perceived to have higher quality. Additionally, products that receive low ratings may face a decrease in sales and, in certain cases, may even become unavailable for purchase. This can result in a smaller customer base and lead to negative reviews. Conversely, positive reviews and product sales could reinforce each other, attracting more buyers who provide favorable reviews and creating a greater number of five-star ratings. The Star Tendency categories help mitigate any potential bias in the rating system. We define the number of reviews with ${Star}^{s}_{i,t}$ of firm $i$ in week $t$ as

\begin{equation}
{Diff({Star}^{s})}_{i,t}=\frac{{{Star}^{s}}_{i,t}}{N_{i,t}}-\frac{{{Star}^{s}}_{i,t-1}}{N_{i,t-1}}
\label{equ:star}
\end{equation}
where $N$ represents the number of reviews published in $n$ weeks, with $n$=1,2\ldots,12. ${Star}^{s}_{i,t}$ represents the rating score equal to $s$, where $s$=1 or 5 for firm $i$. $N_{i,t}$ represents the number of reviews for firm $i$ in week $t$. $Diff({Star}^{s})_{i,t}$ represents the difference between the ${Star}^{s}$ Ratio in week $t$ and week $t-1$. FST and OST variables are complements the explanation of the CNST and CPST variables based on the numerical way. Table \ref{tab:sum} shows that the mean of $Diff({Star}^{1})_{i,t}$ is approximately zero and the interquartile range is 0.02. The mean of $Diff({Star}^{5})_{i,t}$  is also approximately zero, with an interquartile range of 0.05.

\subsubsection{Control Variables}

To generate additional firm-level variables, we acquired financial statement data and stock return data from the China Stock Market and Accounting Research (CSMAR) database. In addition to commonly used firm characteristics such as return on assets(\textit{ROA}), book-to-market (\textit{B/M}), and size (\textit{Size}), we also considered several characteristics likely to be connected to consumer opinions or stock return predictability. To avoid using consumer reviews and ratings as a proxy for a firm’s operating performance inferred from financial statements, we constructed other variables such as idiosyncratic volatility (\textit{Ivol}, \citealt{ang2006cross,boyer2010expected}), asset growth (\textit{AG}, \citealt{cooper2008asset}), illiquidity (\textit{Illiq}, \citealt{amihud2002illiquidity}), market risk (\textit{Beta}, \citealt{frazzini2014betting}), gross profitability (\textit{GP}, \citealt{novy2013other}), and turnover rate (\textit{Turn}, \citealt{jiang2018q}). 
\textit{Ivol} is the weekly standard deviation of daily residuals estimated from the Fama–French three-factor model. 
\textit{AG} is the quarterly growth rate of total assets. 
\textit{Illiq} is calculated as the absolute price change scaled by volume.
\textit{Beta}  is determined as the coefficient obtained by regressing monthly stock returns against market returns using a rolling window of fifty days. We define \textit{GP}as the proportion of pre-extraordinary income to the book value of assets. 
\textit{Turn} is calculated as the mean value of weekly turnover, representing the ratio of shares traded to shares outstanding, over the previous week. Additionally, we took into account expenses on advertising (\textit{Ad}) and research and development (\textit{R\&D}) as these can influence a company's perception among consumers. 
The China Asset Management Research Centre (CAMRC)\footnote{Source: http://sf.cufe.edu.cn/info/1198/12041.htm} provides the excess market returns (\textit{MKT}) as well as the following factors: high-minus-low (\textit{HML}), small-minus-big (\textit{SMB}) and winners-minus-losers (\textit{UMD}).

\subsubsection{Other Variables}

~\\
\textbf{Measuring Sentiment}
~\\

Following \citep{wang2021investor}, we utilize the consumer confidence index (CCIs) as a proxy for sentiment in China’s A-share market, recognizing that there is a positive relationship between consumer confidence and investor sentiment \citep{lemmon2006consumer,qiu2004investor}. The CCIs is an indicator that measures consumer confidence and captures their assessments of the state of the economy today, as well as their subjective perceptions of future prospects, income levels, income expectations, and psychological states related to consumption. It serves as a leading indicator for forecasting consumption and economic developments. CCIs is often seen as an essential addition to total consumption. The CCIs comprises two components: the consumer expectation index and consumer satisfaction index. Moreover, we convert the monthly CCIs into weekly ones to ensure frequency consistency. Table \ref{tab:sum} presents the mean and standard deviation of the CCIs, which are 107.93 and 7.24, respectively.

~\\
\textbf{Measuring Accounting transparency}
~\\

The level of accounting transparency in a firm is determined by the extent to which its reported accounting earnings accurately reflect its genuine economic earnings. To measure accounting transparency, we commonly use proxies such as earnings aggressiveness (EA) and earnings smoothing (ES) \citep{bhattacharya2003world}. Therefore, we use EA and ES measures for the firm’s accounting transparency. 

We deﬁne the EA of ﬁrm $i$ in year $y$ as
\begin{equation}
{EA}_{i,y}=\frac{{ACC}_{i,y}}{{TA}_{i,y-1}}
\label{equ:ea}
\end{equation}

We deﬁne the ES of ﬁrm $i$ in year $y$ as
\begin{equation}
{ES}_{i,y}=\frac{Std\left(\ \frac{{CFO}_{i,y-3}}{A_{i,y-4}},\frac{{CFO}_{i,y-2}}{A_{i,y-3}},\frac{{CFO}_{i,y-1}}{A_{i,y-2}},\frac{{CFO}_{i,y}}{A_{i,y-1}}\right)}{Std\left(\ \frac{{NI}_{i,y-3}}{A_{i,y-4}},\frac{{NI}_{i,y-2}}{A_{i,y-3}},\frac{{NI}_{i,y-1}}{A_{i,y-2}},\frac{{NI}_{i,y}}{A_{i,y-1}}\right)}
\label{equ:es}
\end{equation}

where$ {ACC}_{i,y}$ is the accruals of ﬁrm $i$ in year $y$; ${TA}_{i,y-1}$ is defined as the total assets of firm $i$ in year $y-1$; $Std$ represents the standard deviation in parentheses; ${CFO}_{i,y-k}$ represents the variance in cash flow from operations of firm $i$ in year $y-k$, where $k$=0,1,2,3; ${NI}_{i,y-k}$ defined as net profit for ﬁrm $i$ in year $y-k$, where $k$=0,1,2,3. Moreover, we convert the quarterly EA and ES into weekly ones to ensure frequency consistency.
Table \ref{tab:sum} displays the summary statistics for EA and ES, indicating a mean of 0.13 and 4.13, respectively.

~\\
\textbf{Measuring Cash flow}
~\\

Numerous studies suggest that when customers perceive a superior level of quality, firms can benefit from increased future cash flows. This can be achieved by improving their standing with customers, lowering price elasticity and reducing marketing expenses, increasing the range of available products, and lowering informational costs for consumers \citep[e.g.][]{allen1984reputation,choi1998brand,cabral2000stretching}. The proliferation of internet-based consumer opinion platforms has the potential to amplify these effects, as consumers now play an active role in producing and accessing product-related information. Consumer opinions can also reveal novel information about financial flows. To capture cash flow surprises, we construct two measures.

First, we measure revenue surprises (SUR) using a standardized unexpected revenue growth estimator \citep{jegadeesh2006revenue}. Specifically, we define the SUR of firm $i$ in quarter $q$ as

\begin{equation}
{SUR}_{i,q}=\frac{{REV}_{i,q}-{REV}_{i,q-4}}{\sigma_{i,q}}
\label{equ:sur}
\end{equation}
where ${REV}_{i,q}$ represents the revenue per share for ﬁrm $i$ in quarter $q$, ${REV}_{i,q-4}$ represents the revenue per share for ﬁrm $i$ in quarter $q-4$, and $\sigma_{i,q}$ represents the standard deviation of the growth in revenue for ﬁrm $i$ in quarter $q$. 

The second measure is earnings surprises (SUE), which we quantify using the standardized unexpected earnings estimator \citep{jegadeesh2006revenue}. We deﬁne the SUE of ﬁrm $i$ in quarter $q$ as

\begin{equation}
{SUE}_{i,q}=\frac{{Earnings}_{i,q}-{Earnings}_{i,q-4}}{\sigma_{i,q}}
\label{equ:sue}
\end{equation}
where ${Earnings}_{i,q}$ represents the operating profit for ﬁrm $i$ in quarter $q$, ${Earnings}_{i,q-4}$ represents the operating profit for ﬁrm $i$ in quarter $q-4$, and $\sigma_{i,q}$ represents the standard deviation of the quarterly operating profit for ﬁrm $i$ in quarter $q$. Additionally, we convert the quarterly SUE and SUR into weekly ones to ensure frequency consistency. The mean values for SUR and SUE are presented in Table \ref{tab:sum} and are 0.24 and 0.47, respectively.

To better capture cash flow shocks for firms, we use measures of profitability shock. Previous research has shown that there is a positive and significant correlation between profitability shocks and stock returns \citep{yin2021big}. To construct our measures, we follow the methodologies of \citet{hou2019resurrecting} and \citet{yin2021big}, which involve using both the HVZ and VOL models. In contrast to the HVZ model, the VOL model has been designed to incorporate earnings volatility for the purpose of conducting a more robust analysis. Additionally, we convert the quarterly profitability shocks into weekly ones to ensure frequency consistency.

The HVZ model and the VOL model are defined as follows:

(1) HVZ model
\begin{equation}
{Prof}_{i,q+1}=\alpha_{0,q}+\alpha_{1,q}\frac{V_{i,q}}{A_{i,q}}+\alpha_{2,q}{DD}_{i,q}+\alpha_{3,q}\frac{D_{i,q}}{B_{i,q}}+\alpha_{4,q}{Prof}_{i,q}+\varepsilon_{i,q+1}
\label{equ:hvz}
\end{equation}
where ${Prof}_{i,q}$ is the profitability for firm $i$ in quarter $q$; $\frac{V_{i,q}}{A_{i,q}}$ represents the ratio of an asset’s market value to its book value, and ${DD}_{i,q}$  is a dummy variable with a value of 1 indicating that firm $i$ does not pay dividends in quarter $q$, and a value of 0 indicating that the firm pays dividends in quarter $q$; $\frac{D_{i,q}}{B_{i,q}}$ defined as the dividend payments of firm $i$ in quarter $q$. Profitability in quarter $q$ is measured using ROA.

(2)	VOL model
\begin{equation}
{Prof}_{i,q+1}=\alpha_{0,q}+\alpha_{1,q}\frac{V_{i,q}}{A_{i,q}}+\alpha_{2,q}{DD}_{i,q}+\alpha_{3,q}\frac{D_{i,q}}{B_{i,q}}+\alpha_{4,q}{Prof}_{i,q}+\alpha_{5,q}{Vol}_{i,q}+\varepsilon_{i,q+1}
\label{equ:vol}
\end{equation}
where ${Vol}_{i,q}$ is defined as the earnings volatility of firm $i$ in quarter $q$. he calculations for the remaining variables align with those utilized in the HVZ model. We then compute the expected profitability for quarter $q+1$ by multiplying the independent variables from quarter $q$ by the coefficients derived from the regressions for quarter $q-1$. Subsequently, we define the profitability shock (a proxy for cash flow shock) in quarter $q+1$ as:
\begin{equation}
{Prof Shock}_{q+1}={Realized\ Prof}_{q+1}-{Expected\ Prof}_{q+1}
\label{equ:prof}
\end{equation}
Moreover, we convert the quarterly profitability shock into weekly ones to ensure frequency consistency.

Table \ref{tab:sum} presents the summary statistics of these main variables in this sample, and we report the mean, standard deviation, minimum, 25th percentile, median, 75th percentile, and maximum. Detailed definitions for fall variables are reported in  Appendix \ref{tab:app}.

\centerline{[Insert Table \ref{tab:sum} here]}%

\section{Results}
\label{sec:res}
In this section, we present our empirical findings on the informational value of customer reviews.

\subsection{Customer Online Satisfaction and Stock returns}

Our objective is to investigate the impact of the Customer Sentiment Tendency category on future stock returns in China’s A-share market. To achieve this, we begin by examining the effect using a static panel regression that accounts for fixed effects. Specifically, we run the following panel regression to analyze the effect:

\begin{equation}
{R}_{i,t+1}=\alpha+\tau_{i}+\lambda_{i}+\beta_{1} Diff(\mathrm{rev})^{neg}_{i,t}+\gamma \mathrm{X}_{i,t}+\varepsilon_{i,t+1}
\label{equ:statneg}
\end{equation}
\begin{equation}
{R}_{i, t+1}=\alpha+\tau_{i}+\lambda_{i}+\beta_{1} Diff(\mathrm{rev})^{pos}_{i,t}+\gamma \mathrm{X}_{i,t}+\varepsilon_{i,t+1}
\label{equ:statpos}
\end{equation}

To address endogeneity problems, we run a dynamic panel model using GMM estimators \citep{arellano1991some} and run the following regression:
\begin{equation}
{R}_{i,t+1}=\alpha+\beta_{1} R_{i,t}+\beta_{2} Diff(\mathrm{rev})^{neg}_{i,t}+\gamma \mathrm{X}_{i,t}+\varepsilon_{i,t+1}
\label{equ:dyneg}
\end{equation}

\begin{equation}
{R}_{i,t+1}=\alpha+\beta_{1} R_{i,t}+\beta_{2} Diff(\mathrm{rev})^{pos}_{i,t}+\gamma \mathrm{X}_{i,t}+\varepsilon_{i,t+1}
\label{equ:dypos}
\end{equation}

where ${R}_{i,t}$ and  ${R}_{i,t+1}$ are defined as the weekly returns for firm $i$ in week  $t$ and $t+1$, respectively; $Diff(\mathrm{rev})^{neg}_{i,t}$ and $Diff(\mathrm{rev})^{pos}_{i,t}$ represent the Customer Negative Sentiment Tendency (CNST) and Customer Positive Sentiment Tendency (CPST) for firm $i$ in week $t$; $\mathrm{X}_{i,t}$ is defined as the vector of control variables for firm $i$ in week $t$, including ﬁrm size (\textit{Size}), book-to-market ratio (\textit{B/M}), return on assets (\textit{ROA}), idiosyncratic volatility (\textit{Ivol}, \citealt{ang2006cross,boyer2010expected}), asset growth (\textit{AG}, \citealt{cooper2008asset}), illiquid (\textit{Illiq},\citealt{amihud2002illiquidity}), market risk (\textit{Beta}, \citealt{frazzini2014betting}), gross profitability (\textit{GP}, \citealt{novy2013other}), turnover rate (\textit{Turn},\citealt{jiang2018q}), and expenses on advertising (\textit{Ad})and research and development (\textit{R$\&$D}); $\tau_i$ represents the time fixed effects, and $\lambda_i$ represents the firm fixed effects.                                                                                                    

The findings in Table \ref{tab:cnstcpst} reveal that both static and dynamic models demonstrate a significant negative relationship between CNST and future stock returns. The coefficient for CNST is -0.0008, with t-statistics of -2.7093 and -2.7653 for the static and dynamic models, respectively. The results also suggest a significant economic impact. For example, Model 1 shows that a one standard deviation increase in CNST would result in a significant drop of 0.0008\% in average weekly returns. However, Models 3 and 4 show that CPST is a non-significant positive predictor of future stock returns for both static and dynamic models. This result directly reflects the inverse  relationship between the CNST and stock return. Table \ref{tab:cnstcpst} also reveals several other interesting findings. Consistent with the evidence in \citet{tirunillai2012does}, based on consumer-generated content, positive sentiment words did not predict stock returns, but negative sentiment words negatively predicted stock returns. Moreover, various studies have validated that product recalls have a substantial negative impact on the stock market, indicating that negative consumer feedback on the quality of inferior products can lead to a negative impact on stock returns \citep{hoffer1988impact,barber1996product}.

~\\
\centerline{[Insert Table \ref{tab:cnstcpst} here]}%
~\\

Turning to One-Star Tendency (OST) and Five-Star Tendency (FST), we run the following static and dynamic panel regressions:

\begin{equation}
{R}_{i,t+1}=\alpha+\tau_{i}+\lambda_{i}+\beta_{1} Diff(Star^s)_{i,t}+\gamma \mathrm{X}_{i,t}+\varepsilon_{i,t+1}
\label{equ:statost}
\end{equation}

\begin{equation}
{R}_{i,t+1}=\alpha+\beta_{1} R_{i,t}+\beta_{2} Diff(Star^s)_{i,t}+\gamma \mathrm{X}_{i,t}+\varepsilon_{i,t+1}
\label{equ:dyfst}
\end{equation}

where ${R}_{i,t}$ and ${R}_{i,t+1}$ are the weekly returns for firm $i$ in week $t$ and $t+1$, respectively; $Diff(Star^s)_{i,t}$ represents the One-Star Tendency (OST) and Five-Star Tendency (FST) for firm $i$ in week $t$; $\mathrm{X}_{i,t}$ is defined as the vector of control variables for firm $i$ in week $t$, including \textit{GP}, \textit{Size}, \textit{B/M}, \textit{ROA}, \textit{Illiq}, \textit{Beta}, \textit{Ivol}, \textit{Turn}, \textit{AG}, \textit{Ad} and \textit{R\&D}; $\tau_i$ represents the time fixed effects, and $\lambda_i$ represents the firm fixed effects.

Table \ref{tab:ostfst} presents the regression results for OST and FST, showing a significant negative relationship between OST and future stock returns. In the static model, the coefficient for OST is -0.0072 with a t-statistic of -2.1665, while in the dynamic model, the coefficient for OST is -0.0120 with a t-statistic of -2.6777. In terms of economic magnitude, Model 1 indicates that a one standard deviation increase in OST would result in a statistically significant decline of 0.0072\% (t = -2.1665) in average weekly returns. Consistent with prior work, Table \ref{tab:ostfst} demonstrates that consumers are becoming increasingly significant stakeholders in firms and customers of their products and services due to the rising popularity of internet-based consumer opinion platforms. Customers have access to information about product value and quality and can generate new information about products \citep{huang2015investor,fornell2016stock}. Low ratings indicate that consumers are less satisfied with the product and have more negative attitudes and emotions towards it. However, Models 3 and 4 indicate that FST is not a statistically significant positive predictor of future stock returns for both static and dynamic models. This result also implies that the conclusions of Table \ref{tab:cnstcpst} and Table \ref{tab:ostfst} 's conclusions are mutually validated. For example, the non-significance of FST in both static and dynamic models supports the conclusion from Table \ref{tab:cnstcpst} that CPST is not a significant predictor of future stock returns.

~\\
\centerline{[Insert Table \ref{tab:ostfst} here]}%


\subsection{Sentiment}
Investors have been observed to behave differently under various market conditions \citep{gervais2001learning,nofsinger2005social,li2017investor}. Moreover, we analyze the impact of CNST on future stock returns in different economic settings and use the consumer confidence index (CCIs) as a sentiment proxy in the Chinese stock market \citep{wang2021investor}. To investigate the relationship between sentiment and expected returns, we split the entire sample into two subsamples using the median of CCIs as a cutoff point. Specifically, periods with CCIs larger or smaller than the median of the whole sample are identified as high or low sentiment periods, respectively. We then conduct subsample analyses to examine the effects of different sentiment levels on expected returns. Table \ref{tab:sent} presents the regression results for high and low sentiment periods. The results show that CNST can significantly negatively predict future stock returns during high sentiment periods. However, low sentiment does not significantly predict future stock returns.

Consistent with the prior work, investors tend to buy stocks when they are optimistic and sell them when they are pessimistic \citep{baker2006investor,chung2012does}. Therefore, this implies that during periods of high sentiment, investors are more likely to engage in trading, thereby exerting a stronger impact on stock markets. Our findings are consistent with this pattern, as CNST during high-sentiment periods can negatively predict future stock returns.

~\\
\centerline{[Insert Table \ref{tab:sent} here]}%

\subsection{Growth Companies and Value Companies} 

Consumer review data is likely to contain information related to a company’s future business prospects, especially regarding consumer demand for the company’s products or purchase intentions. Investors may react differently to such information for different types of companies. For growth companies, investors may react more strongly to information about the company’s prospects, as the market valuation of growth companies is more sensitive to their future growth prospects. Moreover, such companies have great development potential, but their future development may be unstable and face more uncertainty and risk in their future prospects than value companies. Therefore, consumer reviews and company reputation become more important for the future development of growth companies. Companies experiencing rapid growth may exhibit high levels of profitability and investment. To analyze the impact of company profitability and investment on our results, we use return on assets (ROA) as a measure of profitability and asset growth (AG) as a measure of investment. We then split the entire sample into two subsamples based on the median of ROA and AG. Specifically, periods with ROA and AG larger or smaller than the median of the whole sample are identified as high or low company profitability and high or low company investment, respectively. To further explore the impact of company characteristics on our results, we split the entire sample into two subsamples based on the median of book-to-market (B/M). Specifically, periods with B/M larger or smaller than the median of the whole sample are identified as value or growth companies, respectively.

The regression results for high or low ROA, AG, and B/M are presented in Table \ref{tab:com}. The results in Table \ref{tab:com} show that CNST can significantly negatively predict future stock returns during periods of high ROA, high AG, and low B/M. By contrast, low ROA, low AG, and high B/M do not significantly predict future stock returns. This also confirms that the effect of consumer evaluation data on stock return prediction is indeed caused by investors’ reactions to information about growth and value companies.

\centerline{[Insert Table \ref{tab:com} here]}%

\subsection{Firm’s Accounting Transparency} 
Consumer review data is a type of non-financial information. In accounting and auditing practice, non-financial information is an important tool for judging whether a company engages in earnings management behavior. For example, \citet{chiu2020using} obtained information about a firm’s sales and potential manipulation of operating income by utilizing Google searches of the firm’s products, specifically through the Search Volume Index (SVI). Additionally, a firm’s accounting transparency will also directly affect investors’ behavior in using information. In this subsection, we show whether a firm’s accounting transparency affects investors’ behavior and further affects the predictive power of consumer reviews. If a firm’s accounting transparency is low, market investors and analysts will rely more on non-financial information to make investment decisions. We use earnings aggressiveness (EA) and earnings smoothness (ES) as proxies for accounting transparency (as discussed in Section \ref{section:var}). A high value of EA indicates that a company is likely to engage in earnings management practices to hide its actual economic earnings, which can reduce accounting transparency. Conversely, a more negative value of ES suggests a greater likelihood of earnings smoothing, which could conceal the correlation between earnings and economic performance and result in lower levels of accounting transparency \citep{bhattacharya2003world,liu2017political}. Therefore, higher EA and ES values indicate lower accounting transparency of the company and lower quality of accounting information. To analyze the impact of EA and ES on expected returns, we split the entire sample into two subsamples using the median of EA and ES as the threshold. The high and low EA/ES groups represent firms with different tendencies towards earnings management and smoothing, respectively. We then conduct subsample analyses to investigate the effects of EA and ES on expected returns.

Table \ref{tab:eaes} presents the regression results for high or low EA and ES. The results in Table \ref{tab:eaes} show that CNST can significantly negatively predict future stock returns during periods of high EA and ES. By contrast, low EA and ES do not significantly predict future stock returns. This finding suggests that consumer opinions will be affected by the earnings opacity of the company. This indicates that the predictive power of consumer reviews, a type of non-financial information, is influenced by the degree to which the company manipulates its financial information. These findings suggest that consumer reviews, as a type of non-financial information, have greater predictive power for companies with lower accounting transparency.

~\\
\centerline{[Insert Table \ref{tab:eaes} here]}%

\subsection{Cash Flow} 
To influence stock returns, consumer online opinions need to bring in new information regarding a company’s cash flows. Thus, to capture new information in cash flows, we utilize SUR and SUE. Following \citet{tetlock2008more,da2011searchfun,froot2017measures,chen2014wisdom}, we run the following dynamic panel regressions to resolve endogeneity issues:

\begin{equation}
{SUR}_{i,t}=\alpha+\beta_{1} SUR_{i,t-1}+\beta_{2} Diff(\mathrm{rev})^{neg}_{i,t}+\gamma \mathrm{X}_{i,t-1}+\varepsilon_{i,t+1}
\label{equ:sur2}
\end{equation}

\begin{equation}
{SUE}_{i,t}=\alpha+\beta_{1} SUE_{i,t-1}+\beta_{2} Diff(\mathrm{rev})^{neg}_{i,t}+\gamma \mathrm{X}_{i,t-1}+\varepsilon_{i,t+1}
\label{equ:sue2}
\end{equation}

where ${SUR}_{i,t}$ is revenue surprises of ﬁrm $i$ in week $t$ (as discussed in Section \ref{section:var}); ${SUE}_{i,t}$ is earnings surprises of ﬁrm $i$ in week $t$ (as discussed in Section \ref{section:var}); $Diff(\mathrm{rev})^{neg}_{i,t}$ represents the CNST of firm $i$ in week $t$ (as discussed in Section \ref{section:var}); $\mathrm{X}_{i,t-1}$ is defined as the vector of control variables for firm  $i$ in week $t-1$, including \textit{GP}, \textit{Size}, \textit{B/M}, \textit{ROA}, \textit{Illiq}, \textit{Beta}, \textit{Ivol}, \textit{Turn}, \textit{AG}, \textit{Ad},and \textit{R\&D}; and $\tau_i $ represents the time fixed effects, and $\lambda_i$ represents the firm fixed effects. We also estimate ${SUR}_{i,t-1}$ and ${SUE}_{i,t-1}$ which are the lagged dependent variables. Since the predictability of CNST is to predict the weekly financial index and fundamental information when the financial statement is not published, the prediction depends on the information of CNST of week $t$ and the financial information before week $t$. Therefore, it is necessary to use $t$ and a control variable of week $t-1$. 

We also consider profitability shocks (a proxy for cash flow shocks) \citep{hou2019resurrecting,yin2021big}. We run the following dynamic panel regression: 

\begin{equation}
{Prof Shock}_{i,t}=\alpha+\beta_{1} Prof Shock_{i,t-1}+\beta_{2} Diff(\mathrm{rev})^{neg}_{i,t}+\gamma \mathrm{X}_{i,t-1}+\varepsilon_{i,t+1}
\label{equ:profshock}
\end{equation}

Table \ref{tab:sursue} demonstrates that an increase of one standard deviation in CNST leads to a statistically significant decrease of 0.0020\% (t = -2.0593) in average weekly returns, and the results show that CNST significantly negatively predicts revenue surprises. This result implies that consumer opinions contain tendentious words that carry new information about cash flows. Table \ref{tab:sursue} also indicates that CNST is a significant negative predictor of earnings surprises. The results also demonstrate that a one standard deviation increase in the CNST variable leads to a statistically significant decrease of 0.0703\% (t = -2.3187) in average weekly returns. After controlling for other determinants of revenue and earnings surprises, the predictability of CNST is obtained. For instance, lagged revenue For instance, lagged revenue surprises strongly and positively predict current revenue and earnings surprises. This result implies that the average analyst does not properly incorporate the information from consumer opinions into her estimates.

~\\
\centerline{[Insert Table \ref{tab:sursue} here]}%
~\\

Table \ref{tab:prof} shows the regression results with profitability shocks as the dependent variable. CNST is a significant negative predictor of profitability shocks with different models.  Table \ref{tab:prof} demonstrates that a one standard deviation increase in CNST is associated with a statistically significant decrease of 0.0002\% (t = -1.8839) and 0.0002\% (t = -1.9999) in average weekly returns, respectively. Consistent with prior results, consumer opinions contain tendentious words that contain novel information about cash flow shocks.

~\\
\centerline{[Insert Table \ref{tab:prof} here]}%

\section{Conclusion}
\label{sec:con}
This study investigates the investment value of consumer opinions in China’s A-share market. We collect a comprehensive dataset of customer reviews from JD.com and construct two variables, Customer Negative Sentiment Tendency (CNST) and One-Star Tendency (OST), to quantify customer online satisfaction. To assess the impact of consumer online satisfaction on stock returns, we construct both static and dynamic panel models and control for a vector of control variables. The results show that CNST and OST are negatively associated with subsequent stock returns.

Our findings indicate that expected stock returns are negatively impacted by CNST and OST, and this effect remains consistent across static and dynamic panel data regressions. This negative impact appears to be concentrated among growth companies. Additionally, CNST can significantly negatively predict future stock returns during high sentiment periods and for firms with low accounting transparency. Our findings support the view that consumer online reviews provide new and unique information regarding cash flows, as demonstrated by the negative relationship between CNST and revenue surprises, earnings surprises, and cash flow shocks.

Our empirical results have some worthwhile implications. The findings presented in this paper underscore the significance of consumers as information generators in China’s A-share market. In contrast to conventional information intermediaries such as financial statements or stock analysts, consumer crowds can offer high-frequency information on a firm’s products and cash flows. Consumer online satisfaction can provide forward-looking information for investors and help market participants understand the current and future development of companies from the perspective of their online consumers.

\bibliographystyle{tfcad}
\bibliography{main}

\begin{appendices}
\section{}
Table \ref{tab:app}

\clearpage
\begin{table}[h]
\centering
\caption{Summary statistics on JD.com reviews for 106 public firms. 
This table reports the summary statistics for a sample of JD.com customer reviews for products of public firms from November 2008 through December 2017. To be included in the sample, a firm must have at least 1000 online reviews and a timespan of more than 12 months.We report the number of reviews, products, and public firms for the full sample.
}
\begin{tabular}{@{}lccc@{}}
\toprule
                        & Number of reviews & Number of products & Number of firms \\ \midrule
Final Sample            & 18,008,415        & 164,715            & 106             \\
Home Appliance Industry & 7,245,982         & 44,059             & 25              \\
Garment Industry        & 5,365,380         & 75,774             & 27              \\
Food Industry           & 4,021,617         & 22,964             & 29              \\
Wine Industry           & 1,375,436         & 21,918             & 25              \\ \bottomrule
\end{tabular}
\label{tab:sumreview}
\end{table}

\begin{landscape}
\begin{table}[h]
\centering
\caption{Summary Statistics of Main Variables. 
This table presents the summary statistics for a sample of firms with customer reviews from November 2008 to December 2017. Descriptive statistics for control variables are also included. All variable definitions are provided in Appendix \ref{tab:app}. 
}

\begin{tabular}{@{}lccccccc@{}}
\label{tab:sum}
\tiny
\toprule
Variable                          & Mean   & Std. Dev. & Min     & 25\%   & Median & 75\%   & Max    \\ \midrule
\multicolumn{8}{l}{\textbf{Customer Reviews}}                                                                        \\
\ \ \ \ \ $Diff(Star^1)_{i,t}$                                & 0.00   & 0.11      & -1.00   & -0.01  & 0.00   & 0.01   & 1.00   \\
\ \ \ \ \ $Diff(Star^5)_{i,t}$                                & 0.00   & 0.18      & -1.00   & -0.03  & 0.00   & 0.03   & 1.00   \\
\ \ \ \ \ $Diff(\mathrm{rev})^{neg}_{i,t}$                                   & 0.00   & 1.08      & -41.00  & 0.00   & 0.00   & 0.00   & 37.00  \\
\ \ \ \ \ $Diff(\mathrm{rev})^{pos}_{i,t}$                                  & 0.01   & 2.10      & -65.00  & 0.00   & 0.00   & 0.00   & 56.00  \\ \midrule
\multicolumn{8}{l}{\textbf{Firm-level characteristics}}                                                              \\ 
\ \ \ \ \ Stock Return(R)                       & 0.00   & 0.06      & -0.19   & -0.03  & 0.00   & 0.03   & 0.23   \\
\ \ \ \ \ Advertising(Ad)                    & 0.15   & 0.64      & -27.73  & 0.09   & 0.15   & 0.23   & 1.23   \\
\ \ \ \ \ Book-to-market (B/M)               & 0.53   & 0.22      & 0.04    & 0.36   & 0.50   & 0.68   & 1.24   \\
\ \ \ \ \ R\&D                              & 1.99   & 1.67      & 0.02    & 0.59   & 1.95   & 3.14   & 31.98  \\
\ \ \ \ \ Return On Assets (ROA)             & 0.02   & 0.02      & -0.44   & 0.01   & 0.01   & 0.03   & 0.21   \\
\ \ \ \ \ LnSize (Size)                      & 22.55  & 1.18      & 20.12   & 21.73  & 22.35  & 23.25  & 26.30  \\
\ \ \ \ \ Asset Growth (AG)                  & 0.08   & 0.21      & -0.46   & -0.02  & 0.04   & 0.11   & 3.78   \\
\ \ \ \ \ Illiquid (Illiq)                   & 0.04   & 0.31      & 0.00    & 0.01   & 0.02   & 0.03   & 37.32  \\
\ \ \ \ \ Market risk (Beta)                 & 1.04   & 0.41      & -0.95   & 0.77   & 1.04   & 1.30   & 3.15   \\
\ \ \ \ \ Idiosyncratic volatility (Ivol)    & 0.04   & 0.03      & 0.00    & 0.03   & 0.04   & 0.05   & 1.73   \\
\ \ \ \ \ Gross profitability (GP)         & 0.34   & 0.19      & -0.01   & 0.20   & 0.32   & 0.44   & 1.57   \\
\ \ \ \ \ Turnover rate (Turn)               & 8.91   & 11.88     & 0.01    & 2.78   & 5.26   & 10.42  & 244.62 \\
\ \ \ \ \ Earnings Aggressiveness (ES)      & -0.12  & 0.13      & -0.61   & -0.20  & -0.11  & -0.02  & 0.30   \\
\ \ \ \ \ Earning Smoothing (EA)            & 4.13   & 7.93      & 0.20    & 1.13   & 2.11   & 3.92   & 89.54  \\ \midrule
\multicolumn{8}{l}{\textbf{Cash ﬂow surprises and Sentiment}}                                                        \\ 
\ \ \ \ \ Consumer Confidence Index (CCIs)   & 107.93 & 7.24      & 97.00   & 103.40 & 105.50 & 112.00 & 124.00 \\
\ \ \ \ \ Earnings surprise (SUE)           & 0.47   & 0.99      & -3.07   & -0.08  & 0.48   & 1.12   & 3.07   \\
\ \ \ \ \ Revenue surprise (SUR)            & 0.24   & 1.08      & -3.47   & -0.28  & 0.25   & 0.85   & 3.41   \\ \midrule
\multicolumn{8}{l}{\textbf{Cash flow shock}}                                                                         \\ 
\ \ \ \ \ HVZ model                        & -0.01  & 0.06      & -1.03   & -0.01  & 0.00   & 0.01   & 0.67   \\
\ \ \ \ \ VOL model                        & -0.01  & 0.04      & -0.52   & -0.01  & 0.00   & 0.01   & 0.65   \\ \bottomrule
\end{tabular}
\end{table}
\end{landscape}

\begin{table}[h]
\centering
\caption{Static and Dynamic Models and CPST/CNST. This table presents the results of static and dynamic panel data regressions of stock returns on Customer Negative Sentiment Tendency (CNST) and Customer Positive Sentiment Tendency (CPST), as well as control variables. All other variables are defined in the Appendix \ref{tab:app}. The sample period is from November 2008 to December 2017. The t-statistic is presented in parentheses (* at the 10\% level; ** at the 5\% level; *** at the 1\% level).}
\begin{tabular}{@{}lcccc@{}}
\toprule
 &
  \begin{tabular}[c]{@{}c@{}}Static model\\ (1)\\ $R_{t+1}$\end{tabular} &
  \begin{tabular}[c]{@{}c@{}}Dynamic model\\ (2)\\ $R_{t+1}$\end{tabular} &
  \begin{tabular}[c]{@{}c@{}}Static model\\ (3)\\ $R_{t+1}$\end{tabular} &
  \begin{tabular}[c]{@{}c@{}}Dynamic model\\ (4)\\ $R_{t+1}$\end{tabular} \\ \midrule
$R_t$                &              & 0.0815**      &              & 0.0811**      \\
                     &              & (2.3976)      &              & (2.3911)      \\
$Diff(\mathrm{rev})^{neg}_{i,t}$ & -0.0008***   & -0.0008***    &              &               \\
                     & (-2.7093)    & (-2.7653)     &              &               \\
$Diff(\mathrm{rev})^{pos}_{i,t}$ &              &               & -0.0000      & -0.0001       \\
                     &              &               & (-0.3207)    & (-0.8108)     \\
$Ad_t$               & 0.0007       & 0.0010***     & 0.0007       & 0.0010***     \\
                     & (1.4115)     & (4.0572)      & (1.4118)     & (4.0539)      \\
$B/M_t$               & -0.0313***   & -0.0193***    & -0.0313***   & -0.0193***    \\
                     & (-8.9579)    & (-5.3224)     & (-8.9539)    & (-5.3196)     \\
$R\&D_t$             & 0.0003       & 0.0007        & 0.0003       & 0.0007        \\
                     & (0.7894)     & (1.4339)      & (0.7859)     & (1.4299)      \\
$ROA_t$              & -0.0220      & -0.0365       & -0.0219      & -0.0366       \\
                     & (-1.0234)    & (-1.1803)     & (-1.0218)    & (-1.1810)     \\
$Size_t$             & 0.0023       & -0.0010       & 0.0023       & -0.0010       \\
                     & (1.1620)     & (-1.5882)     & (1.1626)     & (-1.5908)     \\
$Ivol_t$             & 0.0046       & 0.0066        & 0.0043       & 0.0060        \\
                     & (0.4273)     & (0.8057)      & (0.4052)     & (0.7311)      \\
$GP_t$               & 0.0035       & -0.0092*      & 0.0035       & -0.0092*      \\
                     & (0.7246)     & (-1.8587)     & (0.7237)     & (-1.8592)     \\
$Turn_t$             & -0.0003***   & -0.0010***    & -0.0003***   & -0.0010***    \\
                     & (-8.4255)    & (-5.6579)     & (-8.4250)    & (-5.6614)     \\
$Beta_t$             & 0.0007       & 0.0008        & 0.0007       & 0.0008        \\
                     & (0.7182)     & (0.6926)      & (0.6956)     & (0.6614)      \\
$Illiq_t$            & 0.0033***    & 0.0022        & 0.0033***    & 0.0022        \\
                     & (3.1723)     & (0.6968)      & (3.1754)     & (0.7017)      \\
$AG_t$               & 0.0037*      & 0.0056        & 0.0037*      & 0.0056        \\
                     & (1.7464)     & (1.2810)      & (1.7484)     & (1.2822)      \\
Constant             & -0.0341      & 0.0440***     & -0.0341      & 0.0441***     \\
                     & (-0.7644)    & (3.0063)      & (-0.7644)    & (3.0159)      \\
Number of Code       & 106          & 106           & 106          & 106           \\
Year FE              & YES          &               & YES          &               \\
Code FE              & YES          &               & YES          &               \\
AR (1) test p-value  &              & 0.1630        &              & 0.2630        \\
AR (2) test p-value  &              & 0.2190        &              & 0.2220       \\\bottomrule
\end{tabular}
\label{tab:cnstcpst}
\end{table}

\begin{table}[h]
\centering
\caption{Static and Dynamic Models and OST/FST. This table presents the results of static and dynamic panel data regressions of stock returns on One-Star Tendency (OST) and Five-Star Tendency (FST), as well as control variables. All other variables are defined in the Appendix \ref{tab:app}. The sample period is from November 2008 to December 2017. The t-statistic is presented in parentheses (* at the 10\% level; ** at the 5\% level; *** at the 1\% level).}
\begin{tabular}{@{}lcccc@{}}
\toprule
                     & Static model & Dynamic model & Static model & Dynamic model \\
                     & (1)          & (2)           & (3)          & (4)           \\
                     & $R_{t+1}$    & $R_{t+1}$     & $R_{t+1}$    & $R_{t+1}$     \\ \midrule
$R_t$                &              & 0.0826**      &              & 0.0763*       \\
                     &              & (2.4323)      &              & (1.9122)      \\
$Diff(Star^1)_{i,t}$ & -0.0072**    & -0.0120***    &              &               \\
                     & (-2.1665)    & (-2.6777)     &              &               \\
$Diff(Star^5)_{i,t}$ &              &               & 0.0037*      & 0.0063        \\
                     &              &               & (1.6736)     & (1.6114)      \\
$Ad_t$               & 0.0007       & 0.0010***     & 0.0007       & 0.0010***     \\
                     & (1.4136)     & (4.0760)      & (1.4102)     & (3.6401)      \\
$B/M_t$               & -0.0313***   & -0.0192***    & -0.0313***   & -0.0190***    \\
                     & (-8.9570)    & (-5.3099)     & (-8.9558)    & (-4.5878)     \\
$R\&D_t$             & 0.0003       & 0.0007        & 0.0003       & 0.0006        \\
                     & (0.7768)     & (1.4275)      & (0.7949)     & (0.8783)      \\
$ROA_t$              & -0.0219      & -0.0369       & -0.0217      & -0.0423       \\
                     & (-1.0213)    & (-1.1885)     & (-1.0136)    & (-1.1096)     \\
$Size_t$             & 0.0023       & -0.0010       & 0.0023       & -0.0009       \\
                     & (1.1578)     & (-1.6051)     & (1.1581)     & (-1.2916)     \\
$Ivol_t$             & 0.0045       & 0.0063        & 0.0045       & 0.0084        \\
                     & (0.4202)     & (0.7778)      & (0.4209)     & (0.8618)      \\
$GP_t$               & 0.0035       & -0.0092*      & 0.0035       & -0.0059       \\
                     & (0.7131)     & (-1.8641)     & (0.7262)     & (-1.0885)     \\
$Turn_t$             & -0.0003***   & -0.0010***    & -0.0003***   & -0.0009***    \\
                     & (-8.4316)    & (-5.6656)     & (-8.4314)    & (-5.1209)     \\
$Beta_t$             & 0.0007       & 0.0008        & 0.0007       & 0.0010        \\
                     & (0.6973)     & (0.6835)      & (0.6876)     & (0.7954)      \\
$Illiq_t$            & 0.0033***    & 0.0021        & 0.0033***    & 0.0023        \\
                     & (3.1715)     & (0.6877)      & (3.1441)     & (0.6702)      \\
$AG_t$               & 0.0037*      & 0.0056        & 0.0037*      & 0.0042        \\
                     & (1.7623)     & (1.2871)      & (1.7531)     & (0.8831)      \\
Constant             & -0.0338      & 0.0443***     & -0.0339      & 0.0386**      \\
                     & (-0.7591)    & (3.0268)      & (-0.7602)    & (2.5103)      \\
Number of Code       & 106          & 106           & 106          & 106           \\
Year FE              & YES          &               & YES          &               \\
Code FE              & YES          &               & YES          &               \\
AR (1) test p-value  &              & 0.0000        &              & 0.0000        \\
AR (2) test p-value  &              & 0.1530        &              & 0.2750        \\ \bottomrule
\end{tabular}
\label{tab:ostfst}
\end{table}

\begin{table}[h]

\centering
\caption{CNST and Sentiment. This table presents the results of fixed effects panel data regressions of stock returns on CNST and control variables after controlling for the sentiment proxy, which is the consumer confidence index (CCIs). All other variables are defined in the Appendix \ref{tab:app}. The sample period is from November 2008 to December 2017. The t-statistic is presented in parentheses (* at the 10\% level; ** at the 5\% level; *** at the 1\% level).}

\begin{tabular}{@{}lcc@{}}
\toprule
                     & High sentiment & Low sentiment \\ \cmidrule(l){2-3} 
                     & (1)            & (2)           \\
                     & $R_{t+1}$              &$R_{t+1}$            \\ \midrule
$Diff(\mathrm{rev})^{neg}_{i,t}$& -0.0009***     & 0.0000        \\
                     & (-3.0358)      & (0.0005)      \\
$Ad_t$               & 0.0081         & 0.0005        \\
                     & (0.8915)       & (0.8880)      \\
$B/M_t$               & -0.0408***     & -0.0215***    \\
                     & (-8.5786)      & (-3.3856)     \\
$R\&D_t$             & 0.0011**       & -0.0013       \\
                     & (1.9947)       & (-1.0324)     \\
$ROA_t$              & -0.0308        & -0.0011       \\
                     & (-1.0679)      & (-0.0293)     \\
$Size_t$             & 0.0039         & 0.0007        \\
                     & (1.3532)       & (0.1671)      \\
$Ivol_t$             & -0.0091        & 0.0158        \\
                     & (-0.5988)      & (1.0472)      \\
$GP_t$               & 0.0137*        & -0.0009       \\
                     & (1.9274)       & (-0.1016)     \\
$Turn_t$             & -0.0003***     & -0.0005***    \\
                     & (-5.4054)      & (-7.3912)     \\
$Beta_t$             & 0.0005         & -0.0013       \\
                     & (0.4256)       & (-0.6778)     \\
$Illiq_t$            & 0.0034***      & 0.0004        \\
                     & (3.0964)       & (0.1720)      \\
$AG_t$               & 0.0004         & 0.0080*       \\
                     & (0.1418)       & (1.8608)      \\
Constant             & -0.0702        & 0.0066        \\
                     & (-1.0928)      & (0.0755)      \\
Number of Code       & 106            & 106           \\
Year FE              & YES            & YES           \\
Code FE              & YES            & YES           \\ \bottomrule
\end{tabular}
\label{tab:sent}
\end{table}

\begin{table}[h]
\centering
\caption{CNST with Growth Companies and Value Companies. This table presents the results of fixed effects panel data regressions of stock returns on CNST and control variables after controlling for ROA, AG, and B/M. All other variables are defined in the Appendix \ref{tab:app}. The sample period is from November 2008 to December 2017. The t-statistic is presented in parentheses (* at the 10\% level; ** at the 5\% level; *** at the 1\% level).}
\begin{tabular}{@{}lcccccc@{}}
\toprule
                     & High B/M   & Low B/M    & High ROA   & Low ROA    & High AG    & Low AG     \\ \cmidrule(l){2-7} 
                     & (1)        & (2)       & (3)        & (4)       & (5)        & (6)        \\
                     &$R_{t+1}$          &$R_{t+1}$         &$R_{t+1}$          &$R_{t+1}$          &$R_{t+1}$          &$R_{t+1}$          \\ \midrule
$Diff(\mathrm{rev})^{neg}_{i,t}$ & -0.0008    & -0.0012**  & -0.0011*** & -0.0003    & -0.0011*** & -0.0001    \\
                     & (-1.3116)  & (-2.0561)  & (-2.8556)  & (-0.4733)  & (-3.0857)  & (-0.2362)  \\
$Ad_t$               & 0.0142     & -0.0261    & -0.0172    & 0.0009     & 0.0091     & 0.0005     \\
                     & (0.6847)   & (-1.5720)  & (-1.1016)  & (0.0635)   & (0.7256)   & (1.0135)   \\
$B/M_t$               & -0.0493*** & -0.0498*** & -0.0278*** & -0.0356*** & -0.0328*** & -0.0391*** \\
                     & (-4.7662)  & (-4.1255)  & (-4.3244)  & (-5.3741)  & (-6.0919)  & (-7.3784)  \\
$R\&D_t$             & 0.0002     & 0.0008     & -0.0002    & 0.0001     & 0.0009     & -0.0014    \\
                     & (0.1324)   & (0.5473)   & (-0.1688)  & (0.1429)   & (0.8332)   & (-1.1382)  \\
$ROA_t$              & 0.0031     & -0.0990*   & -0.0366    & 0.0012     & -0.0505    & 0.0032     \\
                     & (0.0591)   & (-1.8956)  & (-0.8333)  & (0.0157)   & (-1.3625)  & (0.0992)   \\
$Size_t$             & 0.0017     & 0.0027     & -0.0042    & 0.0086***  & 0.0011     & -0.0010    \\
                     & (0.3234)   & (0.5886)   & (-1.0495)  & (2.6136)   & (0.3237)   & (-0.2901)  \\
$Ivol_t$             & -0.0198    & 0.0083     & -0.0142    & 0.0122     & -0.0097    & 0.0076     \\
                     & (-0.4611)  & (0.3988)   & (-0.3605)  & (0.9992)   & (-0.4732)  & (0.6017)   \\
$GP_t$               & -0.0295    & 0.0171*    & -0.0079    & 0.0116     & 0.0024     & -0.0252**  \\
                     & (-1.2766)  & (1.6857)   & (-0.8543)  & (1.1972)   & (0.3164)   & (-2.5665)  \\
$Turn_t$             & -0.0007*** & -0.0005*** & -0.0003*** & -0.0004*** & -0.0004*** & -0.0004*** \\
                     & (-7.1194)  & (-6.5724)  & (-3.4652)  & (-6.4090)  & (-4.9481)  & (-7.4335)  \\
$Beta_t$             & -0.0007    & 0.0016     & 0.0006     & 0.0012     & -0.0001    & -0.0004    \\
                     & (-0.3200)  & (0.6979)   & (0.3301)   & (0.7199)   & (-0.0620)  & (-0.2592)  \\
$Illiq_t$            & -0.0082**  & 0.0032**   & 0.0347***  & 0.0019*    & -0.0097    & 0.0007     \\
                     & (-2.3907)  & (2.4248)   & (5.7965)   & (1.7212)   & (-0.4732)  & (0.6344)   \\
$AG_t$               & 0.0046     & 0.0090     & 0.0047     & 0.0029     & 0.0024     & 0.0242**   \\
                     & (1.1734)   & (1.5414)   & (0.9021)   & (0.9813)   & (0.3164)   & (2.5157)   \\
Constant             & 0.0122     & -0.0359    & 0.1195     & -0.1716**  & -0.0004*** & 0.0583     \\
                     & (0.0993)   & (-0.3549)  & (1.3267)   & (-2.3861)  & (-4.9481)  & (0.7661)   \\
Number of Code       & 106        & 106        & 106        & 106        & 106        & 106        \\
Year FE              & YES        & YES        & YES        & YES        & YES        & YES        \\
Code FE              & YES        & YES        & YES        & YES        & YES        & YES        \\ \bottomrule
\end{tabular}
\label{tab:com}
\end{table}

\begin{table}[h]
\centering
\caption{CNST and EA/ES. This table presents the results of fixed effects panel data regressions of stock returns on CNST and control variables after controlling for earnings aggressiveness (EA) and earnings smoothing (ES). All other variables are defined in the Appendix \ref{tab:app}. The sample period is from November 2008 to December 2017. The t-statistic is presented in parentheses (* at the 10\% level; ** at the 5\% level; *** at the 1\% level).}
\begin{tabular}{@{}lcccc@{}}
\toprule
                     & High EA    & Low EA     & High ES    & Low ES     \\ \cmidrule(l){2-5} 
                     & (1)        & (2)        &(3)        &(4)        \\
                     & $R_{t+1}$          & $R_{t+1}$          & $R_{t+1}$          & $R_{t+1}$          \\ \midrule
$Diff(\mathrm{rev})^{neg}_{i,t}$& -0.0015*   & -0.0008    & -0.0013*** & -0.0002    \\
                     & (-1.9359)  & (-1.4606)  & (-2.5970)  & (-0.2685)  \\
$Ad_t$               & -0.0225    & -0.0091    & -0.0117    & 0.0006     \\
                     & (-1.2588)  & (-0.4608)  & (-0.7747)  & (1.0942)   \\
$B/M_t$               & -0.0344*** & -0.0471*** & -0.0485*** & -0.0592*** \\
                     & (-3.4547)  & (-6.2619)  & (-5.5438)  & (-6.1868)  \\
$R\&D_t$             & 0.0004     & 0.0009     & -0.0024*   & 0.0080***  \\
                     & (0.1948)   & (0.6283)   & (-1.8229)  & (2.9805)   \\
$ROA_t$              & -0.0139    & -0.1094*   & -0.0516    & -0.0350    \\
                     & (-0.2729)  & (-1.8595)  & (-0.8581)  & (-0.8750)  \\
$Size_t$             & -0.0002    & 0.0088*    & 0.0042     & 0.0039     \\
                     & (-0.0340)  & (1.8137)   & (0.8561)   & (0.5941)   \\
$Ivol_t$             & -0.0078    & 0.0172     & -0.0294    & 0.0092     \\
                     & (-0.2976)  & (0.6875)   & (-0.6692)  & (0.4795)   \\
$GP_t$               & -0.0013    & 0.0364***  & 0.0022     & 0.0093     \\
                     & (-0.0946)  & (2.8141)   & (0.1950)   & (0.8098)   \\
$Turn_t$             & -0.0004*** & -0.0007*** & -0.0008*** & -0.0008*** \\
                     & (-4.1137)  & (-8.6479)  & (-8.3922)  & (-6.7661)  \\
$Beta_t$             & 0.0015     & -0.0008    & 0.0022     & 0.0015     \\
                     & (0.6018)   & (-0.3882)  & (1.0124)   & (0.7016)   \\
$Illiq_t$            & 0.0214***  & 0.0011     & -0.0094*** & 0.0031**   \\
                     & (2.9556)   & (0.9167)   & (-2.6392)  & (2.5165)   \\
$AG_t$               & 0.0004     & 0.0023     & 0.0068*    & -0.0053    \\
                     & (0.0577)   & (0.5395)   & (1.9266)   & (-0.8419)  \\
Constant             & 0.0314     & -0.1734    & -0.0524    & -0.0701    \\
                     & (0.2664)   & (-1.5975)  & (-0.4857)  & (-0.4694)  \\
Number of Code       & 106        & 106        & 106        & 106        \\
Year FE              & YES        & YES        & YES        & YES        \\
Code FE              & YES        & YES        & YES        & YES        \\ \bottomrule
\end{tabular}
\label{tab:eaes}
\end{table}

\begin{table}[h]
\centering
\caption{CNST and Cash Flow Surprises. This table presents the results of dynamic panel data regressions of revenue and earning surprises on CNST and controls described in Eq.\ref{equ:sur2} and Eq.\ref{equ:sue2}. The dependent variable is the standardized unexpected revenue growth estimator (SUR) or the standardized unexpected earnings (SUE). All other variables are defined in the Appendix \ref{tab:app}. The sample period is from November 2008 to December 2017. The t-statistic is presented in parentheses (* at the 10\% level; ** at the 5\% level; *** at the 1\% level).}
\begin{tabular}{@{}lcc@{}}
\toprule
                     & $SUR_{i,t}$ & $SUE_{i,t}$ \\ \midrule
$SUR_{i,t-1}$        & 0.9562***   &             \\
                     & (180.6413)  &             \\
$SUE_{i,t-1}$        &             & 0.9310***   \\
                     &             & (79.8426)   \\
$Diff(\mathrm{rev})^{neg}_{i,t}$ & -0.0020**   & -0.0703**   \\
                     & (-2.0593)   & (-2.3187)   \\
$Ad_{t-1}$               & 0.0021**    & 0.0069***   \\
                     & (2.2882)    & (2.9164)    \\
$B/M_{t-1}$               & -0.0273*    & -0.0658***  \\
                     & (-1.6854)   & (-3.4444)   \\
$R\&D_{t-1}$             & 0.0046**    & 0.0028      \\
                     & (2.2725)    & (1.2853)    \\
$ROA_{t-1}$              & -0.3091**   & -1.0120***  \\
                     & (-2.1165)   & (-3.1458)   \\
$Size_{t-1}$             & 0.0046*     & 0.0076**    \\
                     & (1.9725)    & (2.3892)    \\
$Ivol_{t-1}$             & 0.0033      & -0.0266     \\
                     & (0.0683)    & (-0.2084)   \\
$GP_{t-1}$               & 0.0387*     & 0.0927***   \\
                     & (1.9268)    & (4.0935)    \\
$Turn_{t-1}$             & -0.0004     & -0.0000     \\
                     & (-0.8996)   & (-0.0944)   \\
$Beta_{t-1}$             & 0.0025      & 0.0088      \\
                     & (0.3403)    & (1.1196)    \\
$Illiq_{t-1}$            & 0.0009      & -0.0041     \\
                     & (0.0936)    & (-1.0683)   \\
$AG_{t-1}$               & -0.0197     & 0.0235      \\
                     & (-0.8551)   & (1.0247)    \\
Constant             & -0.0837*    & -0.1545**   \\
                     & (-1.6889)   & (-2.2830)   \\
AR (1) test p-value  & 0.0000      & 0.0000       \\
AR (2) test p-value  & 0.7120      & 0.8960      \\ \bottomrule
\end{tabular}
\label{tab:sursue}
\end{table}

\begin{table}[h]
\centering
\caption{CNST and Profitability Shocks. This table presents the results of a dynamic panel data regression of profitability shocks on CNST and controls described in Eq.\ref{equ:profshock}. The dependent variable is profitability shocks (HVZ model and VOL model). All other variables are defined in the Appendix \ref{tab:app}. The sample period is from November 2008 to December 2017. The t-statistic is presented in parentheses (* at the 10\% level; ** at the 5\% level; *** at the 1\% level).}

\begin{tabular}{@{}lcc@{}}
\toprule
                     & $Prof\ Shock_{i,t}$ & $Prof\ Shock_{i,t}$) \\
                     & (HVZ Model) & (VOL Model)  \\ \midrule
$Prof\ Shock_{i,t-1}$  (HVZ Model)          & 0.6496***          &                    \\
                                 & (15.2823)          &                    \\
$Prof\ Shock_{i,t-1}$ (VOL Model)            &                    & 0.6237***          \\
                                 &                    & (28.2644)          \\
$Diff(\mathrm{rev})^{neg}_{i,t}$ & -0.0002*           & -0.0002**          \\
                                 & (-1.8839)          & (-1.9999)          \\
$Ad_t$                           & -0.0000            & -0.0002            \\
                                 & (-0.0361)          & (-0.9744)          \\
$B/Mt$                           & 0.0098             & 0.0120**           \\
                                 & (1.5754)           & (2.5026)           \\
$R\&D_t$                         & -0.0000            & -0.0000            \\
                                 & (-0.1402)          & (-0.0454)          \\
$ROA_t$                          & 0.0285             & 0.0368             \\
                                 & (0.7597)           & (0.9377)           \\
$Size_t$                         & -0.0008            & -0.0008            \\
                                 & (-1.1603)          & (-1.4661)          \\
$Ivol_t$                         & -0.0105            & -0.0044            \\
                                 & (-0.6100)          & (-0.3174)          \\
$GP_t$                           & -0.0098**          & -0.0073            \\
                                 & (-2.1079)          & (-1.1111)          \\
$Turn_t$                         & 0.0001             & 0.0000             \\
                                 & (1.4105)           & (1.1997)           \\
$Beta_t$                         & 0.0011             & 0.0012             \\
                                 & (1.0764)           & (1.1674)           \\
$Illiq_t$                        & -0.0055            & -0.0074*           \\
                                 & (-1.3352)          & (-1.6916)          \\
$AG_t$                           & -0.0022            & -0.0021            \\
                                 & (-0.8767)          & (-0.8200)          \\
Constant                         & 0.0137             & 0.0121             \\
                                 & (0.9668)           & (1.0718)           \\
AR (1) test p-value              & 0.0185             & 0.0033             \\
AR (2) test p-value              & 0.3380             & 0.3460             \\ \hline
\end{tabular}
\label{tab:prof}
\end{table}

\begin{table}[h]
 \setcounter{table}{0}
 \renewcommand{\thetable}{A\arabic{table}}
 \centering
 \small
 \caption{Variable definitions.}
 
\begin{tabular}{lp{9cm}}
\toprule
Variable                         & Description                                                                                                                                                                                                          \\ \midrule
$Diff(Star^1)_{i,t}$                               & The One-Star Tendency (OST) is derived from the customer review rating score equal to one, as described in Eq.\ref{equ:neg2}.                                                                                                                                      \\
$Diff(Star^5)_{i,t}$                             & The Five-Star Tendency (FST) is derived from the customer review rating score equal to five, as described in Eq.\ref{equ:pos2}.                                                                                                                                \\
$Diff(\mathrm{rev})^{neg}_{i,t}$                                  & The Customer Negative Sentiment Tendency (CNST) represents unfavorable emotions and lack of recommendations for buying certain products, as described in Eq.\ref{equ:neg}.                                                                                       \\
$Diff(\mathrm{rev})^{pos}_{i,t}$                                 & The Customer Positive Sentiment Tendency (CPST) represents favorable emotions and recommendations for buying certain products, as described in Eq.\ref{equ:pos}.                                                                                            \\
Stock Return (R)                 & Stock Return is the weekly individual stock return without considering cash dividend reinvestment.                                                                                                                        \\
Advertising(Ad)                      & Advertising is the ratio of sales expenses to operating revenue.                                                                                                                                                      \\
Book-to-market (B/M)             & Book-to-market is the book value of common equity divided by the market value of common equity.                                                                                                                      \\
R\&D                             & Research and development  is the ratio of R\&D expenses to operating revenue.                                                                                                                                        \\
Return On Assets (ROA)           & Return On Assets is the ratio of net profit to total assets.                                                                                                                                                             \\
LnSize (Size)                    & LnSize is the log value of the market cap of tradable shares.                                                                                                                                                             \\
Asset Growth (AG)                & The quarterly growth rate of total assets, following  \citet{cooper2008asset}.                                                                                                                                           \\
Illiquid (Illiq)                 & The \citet{amihud2002illiquidity} illiquidity measure, calculated as the absolute price change scaled by the volume.                                                                                                                   \\
Market risk (Beta)               & Market risk is determined as the coefficient obtained by regressing monthly stock returns against market returns using a rolling window of fifty days.                                                                 \\
Idiosyncratic volatility (Ivol)  & Idiosyncratic volatility is the weekly standard deviation of daily residuals estimated from the Fama–French three-factor model.                                                                                       \\
Gross profitability (GP)         & Gross profitability is the proportion of pre-extraordinary income to the book value of assets.                                                                                                                       \\
Turnover rate (Turn)             & Turnover rate is calculated as the mean value of weekly turnover, which represents the ratio of shares traded to shares outstanding, over the previous week.                                                        \\
Earnings Aggressiveness (ES)     & Earnings Aggressiveness measures a firm’s accounting transparency, as described in Eq.\ref{equ:es}..                                                                                                                          \\
Earning Smoothing (EA)           & Earning Smoothing measures a firm’s accounting transparency, as described in Eq.\ref{equ:ea}.                                                                                                                                  \\
Consumer Confidence Index (CCIs) & Consumer Confidence Index is the sentiment proxy in China’s A-share market.                                                                                                                                          \\
Earnings surprise (SUE)          & Earnings surprise is the difference between quarterly operating profit and the previous four quarter operating profit scaled by the standard deviation of quarterly operating profit, as described in Eq.\ref{equ:sue}.            \\
Revenue surprise (SUR)           & Revenue surprise is the difference between actual quarterly revenue per share and the previous four-quarter revenue per share scaled by the standard deviation of quarterly revenue growth, as described in Eq.\ref{equ:sur}. \\
HVZ model and VOL model          & These models refer to two different regression model settings for profitability shock (a proxy for cash flow shock), as described in Eq.\ref{equ:hvz} and Eq.\ref{equ:vol}.                                                               \\ \bottomrule

\end{tabular}
\label{tab:app}
\end{table}
\end{appendices}

\end{document}